\documentclass[prl,aps,twocolumn,showpacs]{revtex4}

\usepackage[dvips]{graphicx}
\usepackage{dcolumn}%
\usepackage{amsmath}%
\usepackage{amsfonts}%
\usepackage{amssymb}
 \usepackage[usenames]{color}
\usepackage{appendix}

\providecommand{\U}[1]{\protect\rule{.1in}{.1in}}
\newcommand{\be}{\begin{equation}}
\newcommand{\ee}{\end{equation}}
\newcommand{\bea}{\begin{eqnarray}}
\newcommand{\eea}{\end{eqnarray}}
\newcommand{\bt} {\begin{tabular}}
\newcommand{\et} {\end{tabular}}
\newcommand{\nn}{ \nonumber}
\newcommand{\ds}{\displaystyle}
\newcommand{\ba} {\begin{array}}
\newcommand{\ea} {\end{array}}
\begin{document}

\title{Thermoelectric properties of Marcus molecular junctions}

\author{  Natalya A. Zimbovskaya {\footnote{Corresponding author: natalya.zimbovskaya@upr.edu}}}

\affiliation
{Department of Physics and Electronics, University of Puerto Rico-Humacao, CUH Station, Humacao, PR 00791, USA}

\begin{abstract}
In the present work we theoretically analyze thermoelectric transport in single-molecule junctions  (SMJ) characterized by strong interactions between electrons on the molecular linkers  and phonons in their nuclear environments where electron hopping between the electrodes and the molecular bridge states predominates in the steady state electron transport. The analysis is based on the modified Marcus theory accounting for the lifetime broadening of the bridge's energy levels. We show that the reorganization processes in the environment accompanying electron transport may significantly affect SMJ thermoelectric properties both within and beyond linear transport regime. Specifically, we study the effect of environmental phonons on the electron conductance, the thermopower and charge current induced by the temperature gradient applied across the system.

\end{abstract}

\date{\today}
\maketitle

\subsection{I. Introduction} 

The research community interest in studies of thermoelectric transport in tailored nanoscale systems was triggered by pioneering works of Hicks and Dresselhaus who predicted that heat-to-electricity conversion in small systems may be dramatically enhanced \cite{1,2}. Presently, these studies represent a well established and rapidly developing research field \cite{3,4,5,6,7,8}. The main building block of a tailored nanoscale system is a junction consisting of a couple of conducting electrodes linked by a bridge. A carbon-based nanostructure, a single/multiple quantum dot or a molecule may be used to bridge electrodes.  In such systems, quantum confinement causes the emergence of sharp features in electron transmission spectra. These spectral features allow for the generation of a large open-circuit voltage as a response to an applied temperature difference between the electrodes thus enhancing the efficiency of energy conversion. 

In this work we concentrate on single molecule junctions (SMJ) where the electrodes are linked by a molecule.
Thermoelectric electron transport through a SMJ is governed by a simultaneous action of electric driving forces and thermal gradients. Their combined effect depends on several factors  such as specifics of the bridge coupling to electrodes \cite{9,10,11,12}, Coulomb interactions between electrons on the bridge \cite{13,14,15,16,17}, the interplay between charge and spin transfer \cite{18,19} and quantum interference effects \cite{15,16,20,21,22,23}. Also, thermoelectric transport may be strongly affected by electron-phonon interactions of different kinds. First, electrons traveling through a SMJ  may interact with molecule's vibrational modes \cite{24,25,26,27,28,29}. Furthermore, transport properties of junctions immersed inside a dielectric solvent may undergo significant changes due to the effect of thermalized phonon modes associated with random motions of solvent molecules. 

When electron-phonon interactions are weak, the effect of phonons may be treated as a perturbation of the ballistic electron transport \cite{7,24,25,26}. Then the electron transport could be described using various techniques including the scattering matrix approach, nonequilibrium Green's functions formalism (NEGF) and rate equations. In the opposite limit of strong interactions between electrons and solvent phonons electron transport could be represented as a sequence of hops  between the electrodes and the molecular bridge states where a traveling electron may be temporarily localized by distorting the nearby solvent. Within this transport regime, NEGF is not suitable to compute transport characteristics even accepting very simple models for the bridge, for the calculations become cumbersome and time consuming. It is much easier to carry on the analysis using Marcus theory \cite{30,31,32} which is based on rate equations with specific expressions for electron transfer rates. This theory was successfully employed to study effects of solvent reorganization on charge and heat transport along the molecules \cite{33,34,35,36,37,38,39}. However, the original Marcus theory has some shortcomings. It does not take into account several potentially important factors such as lifetime broadening of molecular levels and friction accompanying solvent reorganization. Modifications of Marcus theory taking these factors into consideration were recently suggested \cite{40,41,42}.

Thermoelectric properties of nanoscale systems are often measured assuming that a thermal gradient $\Delta T=T_L-T_R$ applied across the system is much smaller than the temperatures of the electrodes $T_{L,R}$ and of the solvent temperature $T$ \cite{7,43}. Then the thermoelectric voltage $V_{th}$ which stops the charge current driven by the temperature gradient is linear in $\Delta T$ and the most important transport characteristics, namely, the electron conductance $G$ and the thermopower $S$  directly characterizing heat-to-electric energy conversion do not depend on the temperature gradient. As $\Delta T$ increases, the system may switch to a nonlinear regime of operation \cite{10,44,45}. In this regime, thermoelectric properties of the system are characterized by $V_{th}$ and the thermocurrent $I_{th}$. The thermocurrent is defined as a difference between the charge current flowing through the biased system in the presence ($\Delta T\neq 0$) and absence ( $\Delta T=0$) of the thermal gradient \cite{44}. These characteristics show strongly nonlinear dependencies on $\Delta T$. Both within and beyond linear transport regime, a SMJ may operate as a nanoscale heat engine \cite{46}.

The thermopower of various nanoscale systems was intensely studied. Some earlier results of these studies are summarized in review articles \cite{6,7,43} and research is still going on \cite{,47,48,49,50,51,52}. However, dissipative phenomena originating from the energy exchange between the traveling electron and the solvent surrounding the molecular bridge are so far ignored in the most of these studies, although these phenomena may strongly affect thermoelectric transport characteristics. It was shown that even a moderate coupling of electrons to solvent phonons may bring noticeable changes in the electron conductance and the thermopower \cite{41}.

The present work is inspired by the results reported in Ref.\cite{41}. We generalize these results by extending analysis to the case of stronger coupling of electrons to the solvent and to a nonlinear transport regime. In Sec.II we show the main equations which are used to get results concerning the solvent phonons effect on the characteristics of thermoelectric electron transport through a SMJ discussed in Sec.III. Conclusions are presented in Sec.IV.

\subsection{II.Main equations}

We analyze thermoelectric transport through a SMJ assuming that the bridge is strongly coupled to the environment created by the solvent phonons and weakly coupled to the electrodes. Following earlier works \cite{32,33,37,41} we treat electron-phonon coupling semiclassically. Within Marcus approach, the steady state electron transport between the left($L$) and right ($R$) electrodes is represented by transitions between the molecular linker states with different charge and described by a set of kinetic (rate) equations for probabilities associated with these states. We simulate the molecular bridge by three states assuming that the molecule is neutral being in the state $|b>$ and charged with a single electron being in the states $|a>$ or $|c>$. States corresponding to a doubly charged molecule are supposed to be inaccessible within the considered bias voltage range. Thus we exclude from consideration quantum interference effects. This simplification is justified, for although  such effects could bring significant changes in the characteristics of a nearly ballistic thermoelectric transport, they are negligible in the considered case when the electron transport is essentially incoherent. We also assume that solvent reorganization processes occur solely when electrons are transferred between the electrodes and the bridge and they are characterized by reorganization energies $\lambda_{\alpha}$ ($\alpha=\{a,c\}$) Charging/discharging of electrodes themselves does not affect the solvent.

Probabilities $P_a$, $P_b$ and $P_c$ for the molecule to be in these states at a certain moment $t$ ($P_a+P_b+P_c=1$) are given by rate equations \cite{33,52}:
\be
\frac{dP_a}{dt}=P_b\cdot k_{ba}-P_a\cdot k_{ab}                   \label{1}
\ee
\be
\frac{dP_b}{dt}=P_a\cdot k_{ab}+P_c\cdot k_{cb}-P_b\cdot\left(k_{ba}+k_{bc}\right)           \label{2}
\ee
\be
\frac{dP_c}{dt}=P_b\cdot k_{bc}-P_c\cdot k_{cb}                              \label{3}
\ee
Here, $k_{\alpha b}=k^{L}_{\alpha b}+k^{R}_{\alpha b}$, $k_{b\alpha}=k^{L}_{b\alpha}+k^{R}_{b\alpha}$. The transfer rates $k^{K}_{\alpha b}$ and $k^{K}_{b\alpha}$ ($K=\{L,R\}$) respectively refer to the removal/injection of an electron from/to the bridge. Excluding one of the charged states (e.g. $|c>$) we reduce the molecular linker to a two states system where one state ($|b>$) is neutral and another one ($|a>$) is charged. Despite its simplicity, modeling of the bridge molecule by two states was used in several works \cite{33,37,41,44}. Within this model, the  probabilities $P_a$ and $P_b$ ($P_a+P_b=1$) obey the equations:
\be
\frac{dP_a}{dt}=P_b\cdot k_{ba}-P_a\cdot k_{ab}; \qquad  \frac{dP_b}{dt}=P_a\cdot k_{ab}-P_b\cdot k_{ba}.         \label{4}
\ee 
The steady state probabilities may be easily found solving Eqs.(\ref{1}) -(\ref{4}). For the three states system we get:
\be
P_{b}=\frac{1}{1+\ds\frac{k_{ba}}{k_{ab}}+\ds\frac{k_{bc}}{k_{cb}}}; \qquad  P_{a}=P_{b}\frac{k_{ba}}{k_{ab}}; \qquad  P_{c}=P_{b}\frac{k_{bc}}{k_{cb}}.            \label{5}                 
\ee
Correspondingly, the steady state probabilities for the two states system are given by:
\be
P_a=\frac{k_{ba}}{k_{ab}+k_{ba}};     \qquad   P_b=\frac{k_{ab}}{k_{ab}+k_{ba}}               \label{6}
\ee
The steady state charge current through the SMJ is given by the expression \cite{33,54}:
\be
\frac{I_{ss}}{e}=\frac{\left(I_1\left(1+\ds\frac{k_{ba}}{k_{ab}}\right)+I_2\left(1+\ds\frac{k_{bc}}{k_{cb}}\right)\right)}{\left(1+\ds\frac{k_{ba}}{k_{ab}}+\ds\frac{k_{bc}}{k_{cb}}\right)}                       \label{7}
\ee
where
\be
I_1=\frac{k^{R}_{ab}k^{L}_{ba}-k^{L}_{ab}k^{R}_{ba}}{k_{ab}+k_{ba}};   \qquad  I_2=\frac{k^{R}_{cb}k^{L}_{bc}-k^{L}_{cb}k^{R}_{bc}}{k_{cb}+k_{bc}}  \label{8}
\ee
Note that $eI_{1,2}$ represent steady state charge currents flowing along the system  when transitions $|a>\rightleftarrows|b>$ or $|c>\rightleftarrows|b>$ are forbidden due to inaccessibility of the relevant states. Thus $eI_{1}$ is the expression for the steady state charge current flowing along the bridge simulated by two states.

The transition rates $k^{K}_{\alpha b}$ and $k^{K}_{b\alpha }$ may be presented as \cite{40,41}:
\be
k_{\alpha b}^K =\frac{1}{\pi\hbar}\int_{-\infty}^\infty d \epsilon \Gamma^{K}_{\alpha}(\epsilon) [1 - f_K (\epsilon)]R^{\alpha}_{-}(\epsilon);                                 \label{9}
\ee
\be
k_{b \alpha}^K =\frac{1}{\pi\hbar} 
\int_{-\infty}^\infty d \epsilon \Gamma^{K}_{\alpha}(\epsilon) f_K ( \epsilon)R^{\alpha}_{+}(\epsilon);         \label{10}
\ee
Here, $\Gamma^{K}_{\alpha}(\epsilon)$ describes coupling of the molecular state $|\alpha>$ to the corresponding electrode ($K=\{L,R\}$), $f_K$ are Fermi distribution functions for the electrodes characterized by the temperatures $T_K$ and chemical potentials $\mu_K$ which depend on the bias voltage provided that the system is biased. In further analysis we adopt the wide band approximation for the electrodes and treat $\Gamma^{K}_{\alpha}$ as constants. We also assume that charged molecular states are symmetrically and equally coupled to the electrodes, that is $\Gamma^{L}_a=\Gamma^{R}_a=\Gamma^{L}_c=\Gamma^{R}_c=\Gamma$. The functions $R^{\alpha}_{\pm}$ have the form: 
\begin{align}
R_{\pm}^{\alpha} = &\mathsf{Re}\sqrt{\frac{\pi\beta}{4\lambda_{\alpha}}}\exp\left(\frac{\beta(\Gamma\mp i(\epsilon_{\alpha}\pm\lambda_{\alpha}-\epsilon))^2)}{4\lambda_{\alpha}}\right) 
\nn\\ &\times
\mathsf{erfc}\left(\sqrt{\frac{\beta}{4\lambda_{\alpha}}}\left(\Gamma\mp i(\epsilon_\alpha \pm\lambda_{\alpha}-\epsilon)\right)\right).   \label{11}
\end{align}
In this expression, $\beta=\ds\frac{1}{kT}$, $\epsilon_{\alpha}=E_{\alpha}-E_b$, ($E_\alpha$, $E_b$ being the energies associated with the charged $|\alpha>$ and neutral $|b>$ states of the molecular bridge) and $\mathsf{ erfc(z)}$ is the complimentary error function. As follows from Eq.(\ref{11}), the lifetime broadening of molecular levels is introduced into the expressions for the transfer  rates. One should use this generalization of the original Marcus-Hush-Sidney theory \cite{30,31,32,33} to properly describe thermoelectric transport through single-molecule junctions for the MHS theory gives unphysical results for thermoelectric characteristics of such systems \cite{38,41}.

Assuming that both voltage drop $\Delta V$ and thermal gradient $\Delta T$ applied across the SMJ are small, we can consider thermoelectric transport within the linear transport regime. Accordingly, the charge current shows linear dependence on both $\Delta V$ and $\Delta T$ \cite{54}: $I_{ss}=G\Delta V+G_{th}\Delta T$ where the zero bias electron conductance $G=\ds\frac{\partial I_{ss}}{\partial V}\bigg|_{\Delta V=\Delta T=0}$ and $G_{th}=\ds\frac{\partial I_{ss}}{\partial T}\bigg|_{\Delta V=\Delta T=0}$. Then the thermovoltage $ V_{th}$ required to stop the charge current is proportional to the thermal gradient: $V_{th}=S\Delta T$ where the thermopower $S=-\ds\frac{G_{th}}{G}$. Straightforward calculations basing on Eqs.(\ref{7})-(\ref{10}) bring the following results for $G$ and $S$:
\be
G=G_0\frac{\Gamma}{2}\sum_{\alpha}\left(P_b+P_{\alpha}\right)\frac{k_{\alpha b}L^{(0)}_{\alpha,+}+k_{b\alpha}L^{(0)}_{\alpha,-}}{k_{\alpha b}+k_{b\alpha}} \label{12}
\ee
\be
G_{th}=\frac{e}{\pi\hbar T}\frac{\Gamma}{2}\sum_{\alpha}\left(P_b+P_{\alpha}\right)\frac{k_{\alpha b}L^{(1)}_{\alpha,+}+k_{b\alpha}L^{(1)}_{\alpha,-}}{k_{\alpha b}+k_{b\alpha}}           \label{13}
\ee
where:
\be
L^{(n)}_{\alpha,\pm}=-\int_{-\infty}^{\infty}d \epsilon\frac{\partial f}{\partial\epsilon}(\epsilon-\mu)^{n}R_{\pm}^{\alpha}(\epsilon),     \label{14}
\ee
In Eq.(\ref{14}), the Fermi function is computed supposing that $T_L=T_R=T$ and chemical potential of electrodes in the unbiased junction equals $\mu$. 
When the bridge is simulated by two states, the expressions for $G$ and $S$ are simplified. Instead of Eqs.(\ref{13})-(\ref{14}) we get:
\be
G=G_0\frac{\Gamma}{2}\frac{k_{ab}L^{(0)}_{a,+}+k_{ba}L^{(0)}_{a,-}}{k_{ab}+k_{ba}};           \label{15}
\ee
\be
 S=\frac{1}{T|e|}\frac{k_{ab}L^{(1)}_{a,+}+k_{ba}L^{(1)}_{a,-}}{k_{ab}L^{(0)}_{a,+}+k_{ba}L^{(0)}_{a,-}}             \label{16}
\ee
These results agree with the corresponding results reported in Ref.\cite{41}.

As the temperature difference increases, the system switches to the nonlinear regime of operation. Nonlinear thermoelectric transport in nanoscale systems had been studied in numerous works but the effect of solvent phonons on the transport characteristics was not thoroughly analyzed so far. Here, we analyze this effect by studying the thermally induced contribution to the charge current which is defined as \cite{8,44}:
\be
I_{th}=I_{ss}(V,T_L,\Delta T)-I_{ss}(V,T_L,\Delta T=0).                                  \label{17}
\ee
When a SMJ is immersed in a dielectric solvent, $I_{th}$ may be strongly influenced by nuclear motions in the solvent, as we show below.

\subsection{III.  Results and discussion.}

\subsubsection{(a) Linear transport regime}
\begin{figure}[t] 
\begin{center}
 \includegraphics[width=4.2 cm,height=3.5cm]{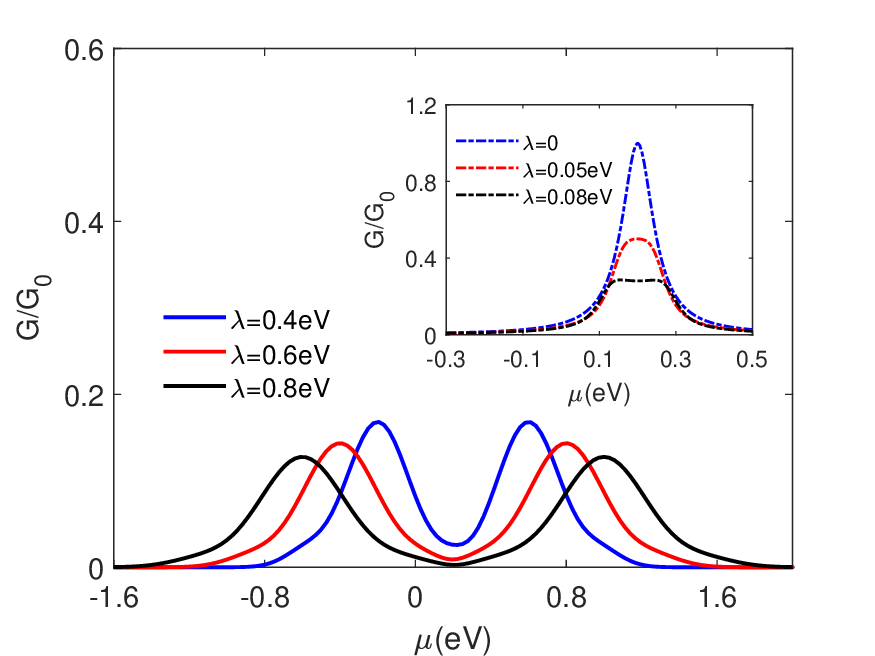}
\includegraphics[width=4.2 cm,height=3.5cm]{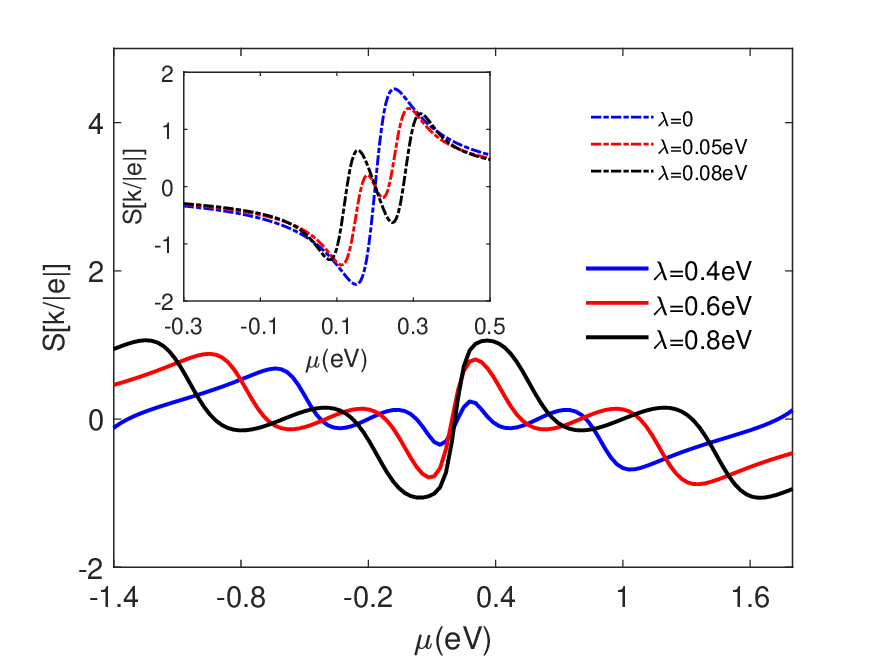}
\caption{Zero bias electron conductance $G$ (left) and the thermopower $S$ (right) as functions of the chemical potential of electrodes $\mu$ computed assuming that the bridge is simulated by two states. Curves are plotted at  the thermal energy $kT=0.026$eV, $\Gamma=0.05$eV, $\epsilon_a=0.2$eV for strong (main bodies) and weak (insets) interactions of an electron on the bridge with the solvent phonons.
}
\label{rateI}
\end{center}\end{figure}
First we consider the behavior of the zero bias electron conductance and the thermopower assuming that the bridging molecule is represented by two states (neutral $|b>$ and charged $|a>$) which is illustrated in Fig.1. If the effect of solvent on the electron transport disappears, $G$ shows a single peak emerging when the electrodes chemical potential $\mu$ crosses the charged molecule energy level with the energy $\epsilon_a$. As for the thermopower $S$, it takes on nonzero values when $\epsilon_a$ is sufficiently close to $\mu$. At $\mu<\epsilon_a$ the state $|a>$ works as LUMO in practical molecules, and charge carriers which travel between the electrodes are mostly electrons. When $\mu$ exceeds $\epsilon_a$ the state $|a>$ starts to serve as HOMO, and the charge carriers pushed through the system by the temperature difference are mostly holes. Thus the thermopower changes its sign at $\epsilon_a=\mu$.

At weak electron-phonon interactions attested by small values of the reorganization energy $\lambda$ the conductance decreases as $\lambda$ enhances, in agreement with the earlier reported results \cite{41}. The thermopower shows more complex behavior. In addition to the decrease of the magnitude accompanying the enhancement of the reorganization energy, $S$ acquires two extra changes of sign occurring at $\mu=\epsilon_a\pm\lambda$. This happens due to the effect of solvent phonons. At stronger electron-phonon coupling the conductance peak splits in two lower but broader maxima situated at $\mu=\epsilon_a\pm\lambda$. At sufficiently large $\lambda$, the conductance becomes negligible over a rather broad interval centered at $\mu=\epsilon_a$. This is the result of Franck-Condon blockade \cite{27,55}. Accordingly, $S$ vs $\mu$ curves assume more complex profiles displayed in Fig.1. Note that at strong interaction between electrons and the solvent phonons the increase of $\lambda$ leads to the enhancement of $S$ magnitude around $\mu=\epsilon_a$. This rather unexpected result originates from the suppression of the conductance due to the Franck-Condon blockade which becomes more pronounced at greater $\lambda$.
\begin{figure}[t] 
\begin{center}
 \includegraphics[width=4.2 cm,height=3.5cm]{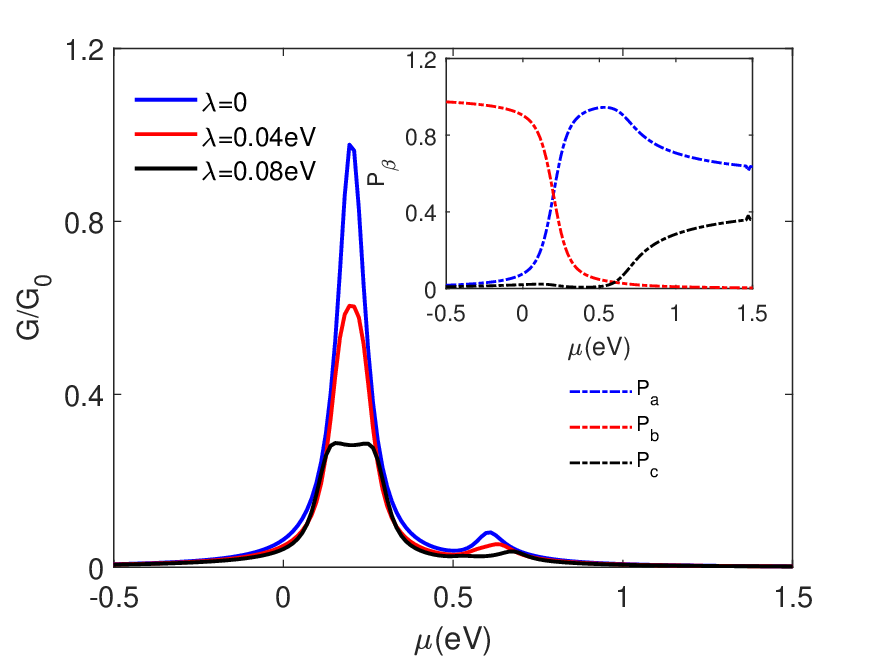}
\includegraphics[width=4.2 cm,height=3.5cm]{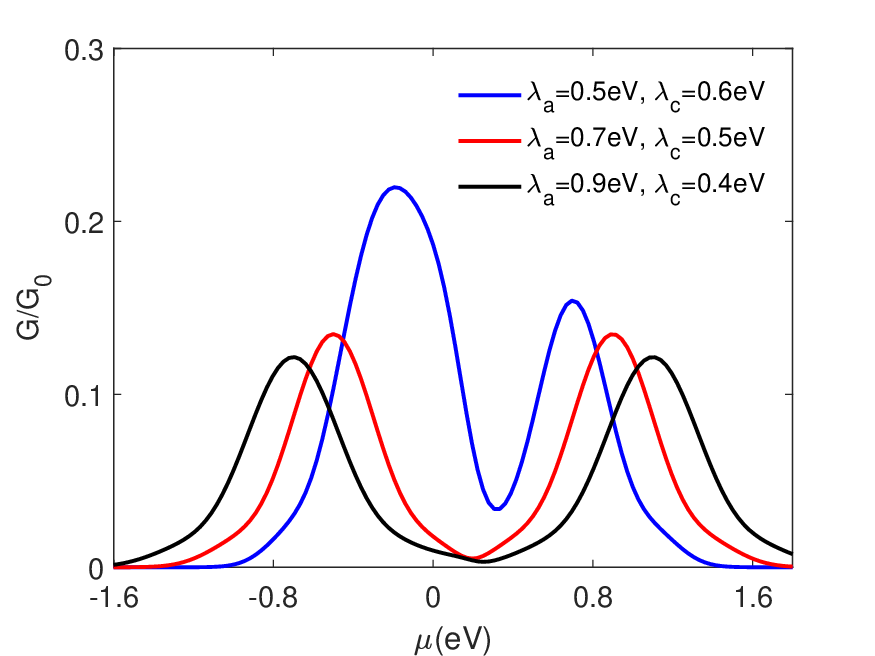}
\caption{Dependencies of zero bias conductance $G$ on the chemical potential $\mu$ in the case when the molecular bridge is treated as a three states system for a weak (left) and strong (right) effect of solvent phonons on the electron transport. Curves are plotted  assuming that $kT=0.026$eV, $\Gamma=0.05$eV, $\epsilon_a=0.2$eV, $\epsilon_c=0.6$eV. Inset shows the steady state occupation probabilities for a three state molecule with the neutral state $|b>$ and charged states $|a>$ and $|c>$ plotted disregarding the effect of solvent phonons at $\epsilon_a=0.2$eV, $\epsilon_c=0.6$eV. 
}
\label{rateI}
\end{center}\end{figure}

Starting to analyze the thermoelectric properties of a SMJ where the bridge is modeled by a three states system, we note that contributions of the charged states into the zero bias electron conductance strongly differ. Assuming  for certainty that $\epsilon_a<\epsilon_c$ we observe that the state $|c>$ contributes to the conductance  significantly less than $|a>$, as shown in the left panel of Fig.2. This difference appears because all transitions corresponding to the injection/removing to/from the molecule more than one electron were omitted from consideration. As a result, the lower state is occupied with a significantly greater probability than the upper one which is illustrated in the inset.

Again, in the case of low intensity of reorganization processes in the solvent, the electron-phonon interactions lead to lowering of the conduction peaks. At stronger coupling of the electron on the bridge to the solvent phonons, the effect of the higher charged state remains nearly negligible provided that the energies $\epsilon_a\pm\lambda_a$ and $\epsilon_c\pm\lambda_c$ are different, as shown in Fig.2. However, if some energy associated with state $|c>$ is sufficiently close to the one associated with $|a>$, the effect of the higher state becomes noticeable. It brings an asymmetry into the double-humped conductance curve and shifts its minimum from the position at $\mu=\epsilon_a$. An example is displayed in the right panel of Fig.2.
\begin{figure}[t] 
\begin{center}
 \includegraphics[width=4.2cm,height=3.5cm]{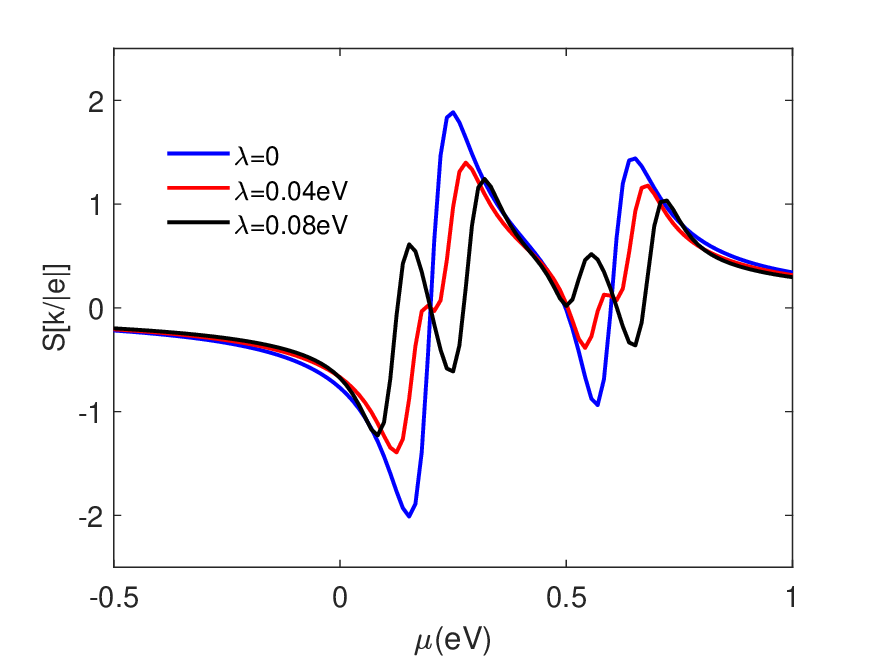}
\includegraphics[width=4.2cm,height=3.5cm]{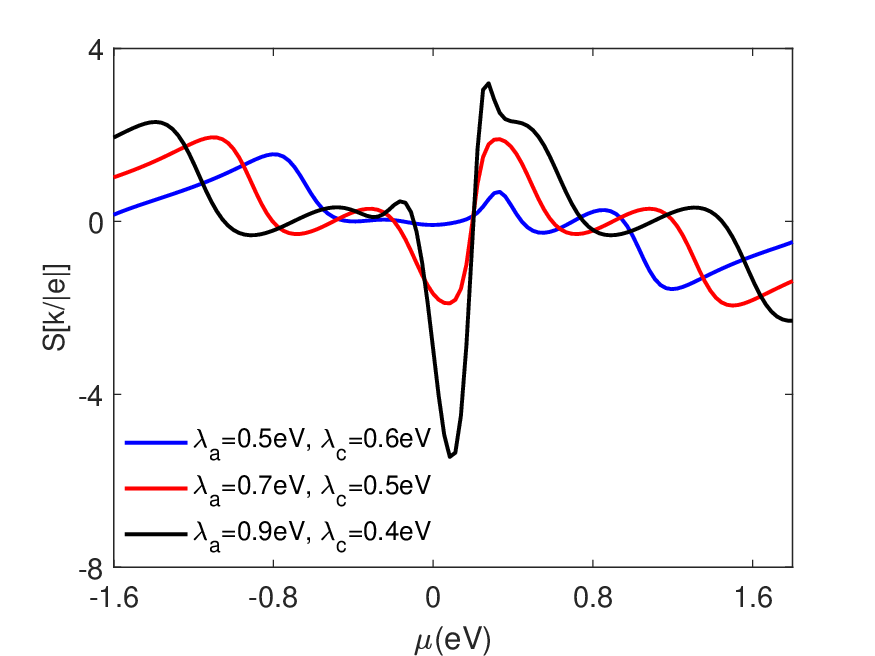}
\caption{ The  thermopower $S$ as a function of $\mu$ for the three states model of the molecular bridge at weak (left) and strong (right) effects of the solvent reorganization on the electron transport. Curves are plotted at $kT=0.026$eV, $\Gamma=0.05$eV, $\epsilon_a=0.2$eV, $\epsilon_c=0.6$eV for several values of reorganization energies.
} 
\label{rateI}
\end{center}\end{figure}

Unlike zero bias conductance, the thermopower is considerably affected  by the influence of a higher state $|c>$. This is especially clearly seen when the energies of the charged states significantly differ from one another and the effects of the solvent reorganization are weak ($\lambda_{\alpha}\ll\epsilon_{\alpha}$), so that the contributions from the two charged states are separated, as shown in Fig.3 (left panel). Note that in the case of $S$, the contribution from the higher energy state is comparable with that coming from the lower state. As the processes of reorganization in the solvent accompanying electron transport intensify, the contributions from the charged states into $S$ overlap and are not easily distinguishable.This is illustrated in the right panel of Fig.3.

In general, the simulation of the bridge by the three states brings results for $S$ which do not contradict those obtained using the two states model considered before. The reorganization processes in the solvent moderately reduce the thermopower magnitude when these processes are not intensive. When the electron-phonon coupling strengthens, especially for the electron in the lower state $|a>$, the thermopower magnitude increases. It may happen that $S$ shows a peak near $\mu=\epsilon_a$ whose height exceeds the maximum thermopower in the absence of the phonons. Again, this may occur because the the conductance around $\mu=\epsilon_a$ at sufficiently large $\lambda_a$ is very close to zero being suppressed by Franck-Condon blockade.
\begin{figure}[t] 
\begin{center}
\includegraphics[width=4.2cm,height=3.5cm]{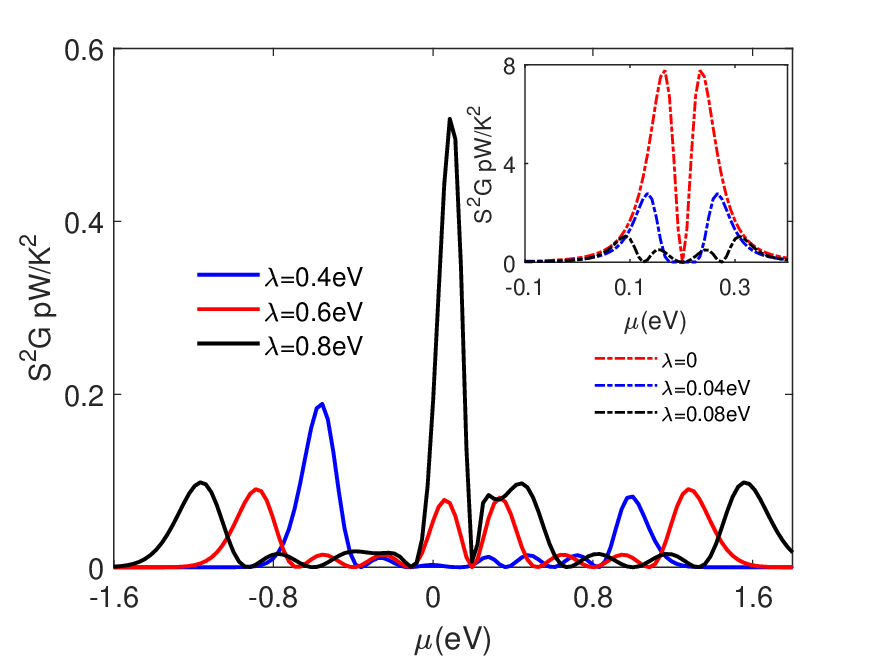}
\includegraphics[width=4.2cm,height=3.5cm]{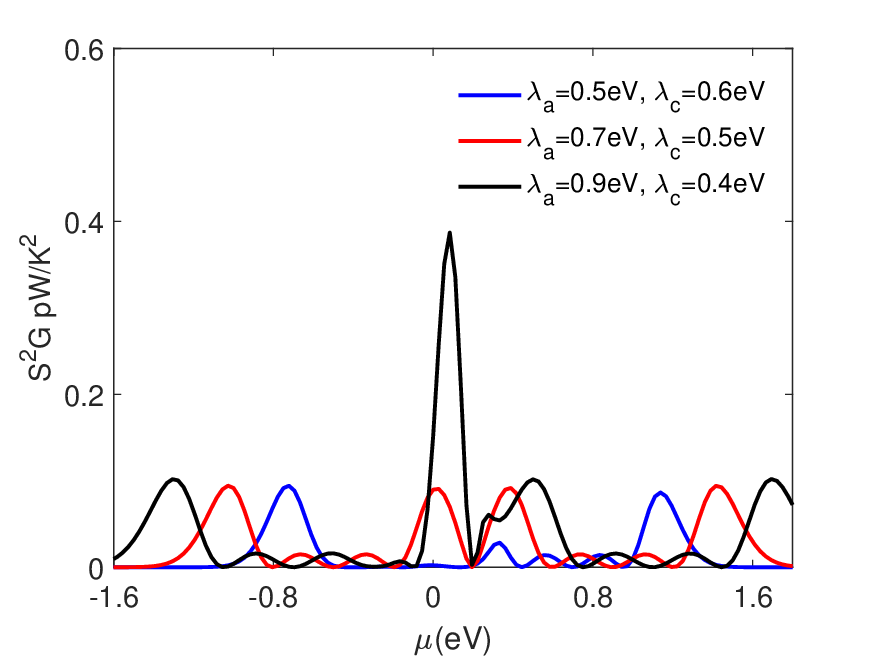}
\caption{ The power factor's $S^2G$ dependencies on $\mu$ plotted assuming that the molecular bridge is represented by the two states (left) and three states (right) assuming that $kT=0.026$eV, $\Gamma=0.05$eV for several values of the reorganization energies. The states energies are: $\epsilon_a=0.2$eV (left) and $\epsilon_a=0.2$eV, $\epsilon_c=0.6$eV (right).
}
 \label{rateI}
\end{center}\end{figure}
The efficiency of practical realization of nanoscale thermoelectric devices may be characterized by the power factor $S^2G$. The behavior of this factor is shown in Fig.4. It appears that the reorganization processes in the solvent  strongly reduce $S^2G$. The same conclusion was already made earlier \cite{41}. The power factor mostly drops due to the fall of the conductace in the region where the thermopower (which is less sensitive to the effect of solvent phonons) reaches its maximum magnitude. It was suggested  that the power factor could be raised by increasing $G$ \cite{41}. This may be achieved either by reducing the electron-phonon coupling strength or by choosing a molecule with the degenerated charged level/levels as the linker in a SMJ. 

\subsubsection{(b) Beyond the linear transport regime: thermocurrent}

 The thermally induced current $I_{th}$ is an important characteristics of thermoelectric transport closely related to the thermovoltage, for $V_{th}$ is the very voltage required to stop $I_{th}$. Correspondingly, similar information concerning the thermoelectric properties of diverse systems may be extracted studying the behavior of either quantity.
As mentioned above, both thermovoltage $ V_{th}$ and thermally induced current $I_{th}$ become nonlinear functions of the temperature gradient at sufficiently larges difference between electrodes temperatures. Temperature dependencies of these important characteristics of nonlinear thermoelectric transport where discussed in numerous works \cite{7,8,26,44,55,56,57}. The research was mostly focused on the thermovoltage behavior, for $V_{th}$ is more directly related to the potential efficiency of the energy conversion. In some works it was suggested  that at sufficiently large $\Delta T$  the efficiency of a nanoscale converter may be enhanced beyond that expected within the linear transport regime \cite{58,59}. However, one meets with significant difficulties trying to achieve this result, for the thermovoltage usually shows a nonmonotonic dependence on the temperature gradient. First increasing in magnitude as $\Delta T$ increases, it may approach zero and then change its polarity at further $\Delta T$ enhancement thus indicating the change of charge carriers (electrons/holes) mostly participating in the transport.

The thermocurrent behavior in nanoscale systems is less thoroughly studied so far. Therefore we choose $I_{th}$ as an object of further research. In accordance with the main purpose of the present work, we analyze how the processes of solvent reorganization may affect the thermocurrent dependencies on the temperature gradient and the bias voltage applied to a SMJ both within and beyond the linear response regime. The analysis is simplified by the fact that the $I_{th}$ definition given by Eq.(\ref{17}) coud be employed to compute the thermocurrent over a whole range of appropriate $\Delta T$ values. For simplicity, we assume that temperature difference is symmetrically distributed between the electrodes: $T_{L,R}=T\pm\ds\frac{1}{2}\Delta T$, $T$ being the solvent temperature.

First we consider $I_{th}$ flowing through an unbiased SMJ where a sole driving force originates from the thermal gradient. We start from a simpler model simulating the bridge by two states. In such a system, the charge current appears when one of the energy levels associated with the charged state $|a>$ and split by the effect of solvent phonons ($\epsilon_a\rightarrow\epsilon_a\pm\lambda$) is close to the chemical potential $\mu$. Then a channel for the charge carriers transport opens up and the charge carriers move to the cooler electrode. The $I_{th}$ direction  at a given $\Delta T$ is controlled by the position of the relevant energy level with respect to $\mu$ determining which kind of charge carriers (electrons/holes) predominates in the transport. The current vanishes if  $\epsilon_a\pm\lambda$ equals $\mu$ for in this case electron and hole fluxes counterbalance each other. The strengthening of electron- phonon coupling leads to reducing of $I_{th}$ magnitude because fluctuations of the solvent electric potential in the vicinity of the bridge accompanying the reorganization processes are hindering the orderly charge flow along the bridge molecule. The $I_{th}$ behavior for the case of the two states model for the bridge is shown in Fig.5 (see inset). Note that the thermocurrent approaches zero at large values of $|\Delta T|$ when neither transport channel may be open since neither relevant energy could be sufficiently close to $\mu$.

Simulating the bridge by a three states system complicates the analysis for in this case transport channels associated with both charged states may contribute to the charge transfer. Accordingly, the thermocurrent dependencies on the temperature gradient vary following alternations in $\epsilon_\alpha$ and $\lambda_\alpha$, as shown in Fig.5. Nevertheless, some basic features in the thermocurrent behavior remain unchanged, namely, its direction reversal at $\Delta T=0$, its tendency to approach zero at large temperature differences between the electrodes and its linear dependence on $\Delta T$ at small values of $|\Delta T|$ corresponding to the linear transport regime.
\begin{figure}[t] 
\begin{center}
\includegraphics[width=4.2cm,height=3.5cm]{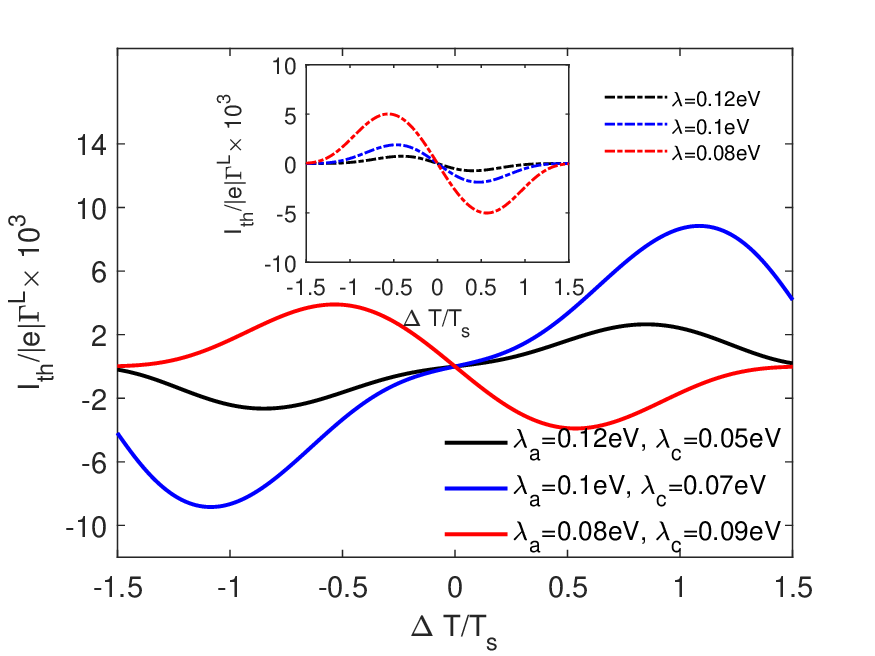} 
\includegraphics[width=4.2cm,height=3.5cm]{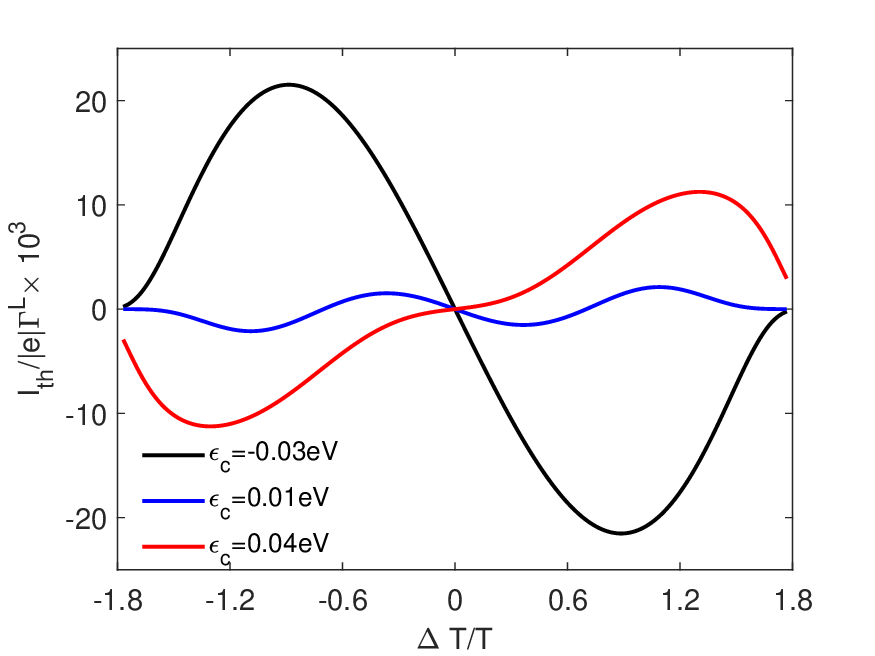}
\caption{Thermally induced current $I_{th}$ flowing through an unbiased SMJ. The bridge is simulated by a three states system with two charged states $|a>$  and $|c>$. Curves are plotted assuming the thermal energy $kT=0.026$eV, $\Gamma=0.01$eV, $\epsilon_a=-0.02$eV at fixed value of $\epsilon_c=0.03$eV and several values of the reorganization energies (left) and at fixed values of reorganization energies ($\lambda_a=0.1$eV, $\lambda_c=0.07$eV) and several values of $\epsilon_c$ (right). The inset shows $I_{th}$ vs $\Delta T$ computed for a two states bridge at $\epsilon_a=-0.02$eV and several values of the reorganization energy.
}
 \label{rateI}
\end{center}\end{figure}

When the SMJ is biased, the simultaneous action of thermal and electric driving forces shows at moderate magnitudes of the bias voltage $V$. In strongly biased molecular junctions the effect of the temperature gradient becomes negligible compared to that of the bias voltage, and $I_{th}$ vanishes, as follows from Eq.(\ref{17}). In a biased system, the chemical potentials of electrodes differ and their values are governed by the bias voltage distribution. Here, we consider the case of a symmetrically distributed voltage assuming that $\mu_{L,R}=\mu\pm\ds\frac{1}{2}V$. The difference between the electrodes chemical potentials creates the conduction window whose boundaries are somewhat blurred because of the effect of the electrodes temperatures. The blur is more pronounced for the boundary associated with the hot electrode and less for that associated with the cold one. 
\begin{figure}[t] 
\begin{center}
\includegraphics[width=4.2cm,height=3.5cm]{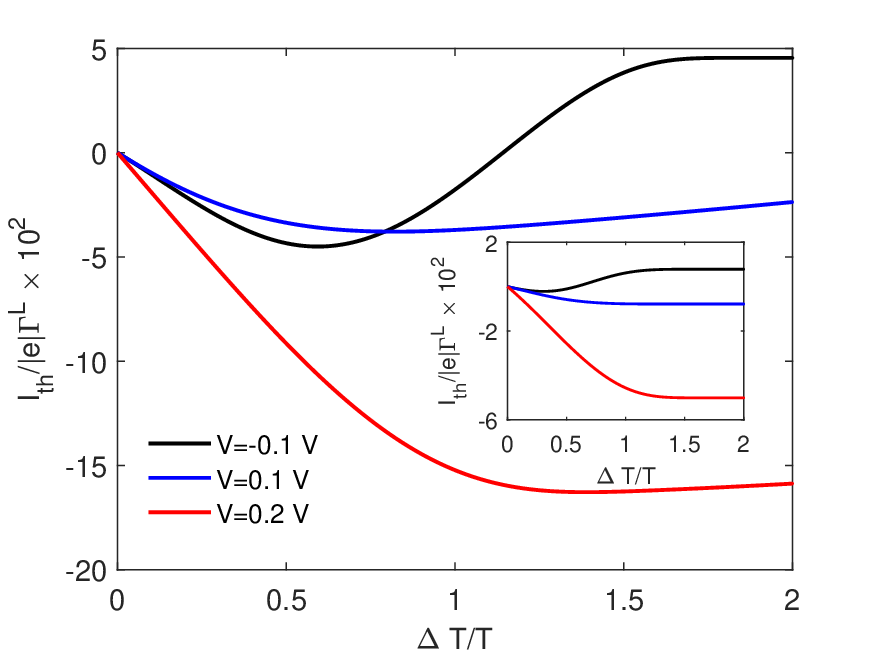} 
\includegraphics[width=4.2cm,height=3.5cm]{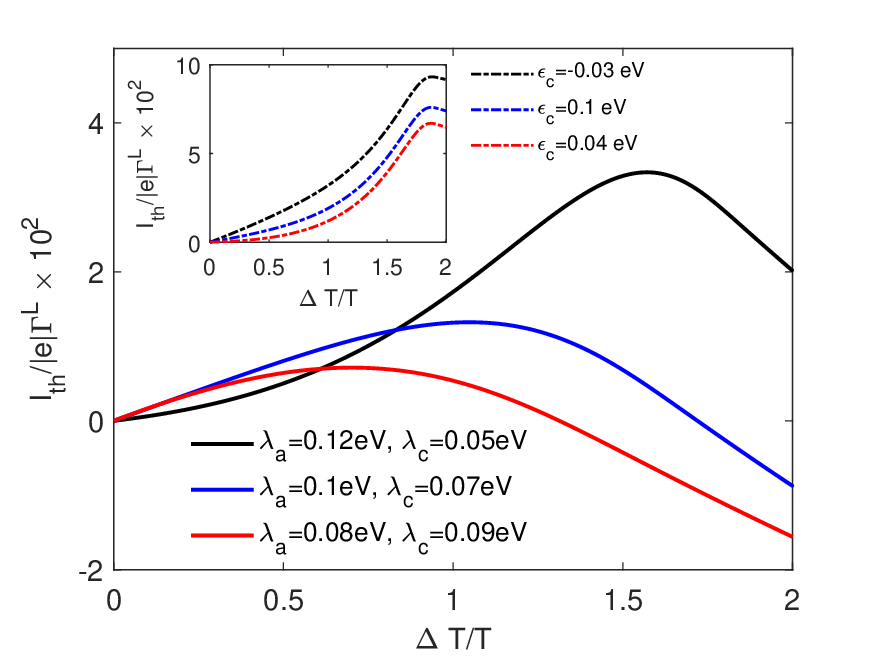}
\caption{Thermally induced current $I_{th}$ flowing through a biased SMJ. The bridge is simulated by a three states system with two charged states $|a>$  and $|c>$. Curves are plotted assuming the thermal energy $kT=0.026$eV, $\Gamma=0.01$eV, $\epsilon_a=-0.02$eV, $\epsilon_c=0.03$eV, $\lambda_a=0.15$eV, $\lambda_c=0.05$eV and several values of the bias voltage (left) and at a fixed value of the bias voltage $V=0.3$V and several values of the reorganization energies (right).  Left inset shows $I_{th}$ vs $\Delta T$ curves plotted for the two states system assuming $\epsilon_a=-0.02$eV, $\lambda_a=0.15$eV for several values of $V$. Right inset shows $I_{th}$ vs $\Delta T$ curves plotted at $V=0.3$V, $\epsilon_a=-0.02$eV, $\lambda_a=0.12$eV, $\lambda_c=0.05$eV at several values of $\epsilon_c$.
}
 \label{rateI}
\end{center}\end{figure}
As discussed above, the effect of solvent reorganization leads to splitting of each peak in the conductance in two centered at the energies $\epsilon^{\pm}_\alpha=\epsilon_\alpha\pm\lambda_\alpha$. Simplifying the matter, we may treat these energies as conditional 'energy levels' indicating possible transport channels. A transport channel opens up when a boundary of the conduction window approaches the energy $\epsilon^{\pm}_\alpha$. When the boundary crosses each of this 'levels'  the thermocurrent changes its direction. Thus positions of the charged bridge levels and intensity of reorganization processes in the solvent may strongly affect $I_{th}$. This is illustrated in Fig.6. Again, $I_{th}$ is small and its magnitude is proportional to $\Delta T$ when the latter is much smaller than the electrodes temperatures. At larger $\Delta T$ the thermocurrent shows diverse behaviors determined by the bias and the the positions of $\epsilon_{\alpha}^{\pm}$ with respect to the boundaries of the conduction window. In general, beyond the linear response regime the thermocurrent magnitude enhances. Nevertheless, $I_{th}$ may approach zero and then change its direction at certain $\Delta T$. As in the case of $V_{th}$, this indicates  the change of the predominant charge carriers.  As in the case of $V_{th}$, this indicates  the change of the predominant charge carriers and makes it difficult to estimate the potential thermoelectric performance of the system beyond the linear transport regime

Dependencies of $I_{th}$ on the bias voltage $V$ are presented in Fig.7.

\begin{figure}[t] 
\begin{center}
\includegraphics[width=4.2cm,height=3.5cm]{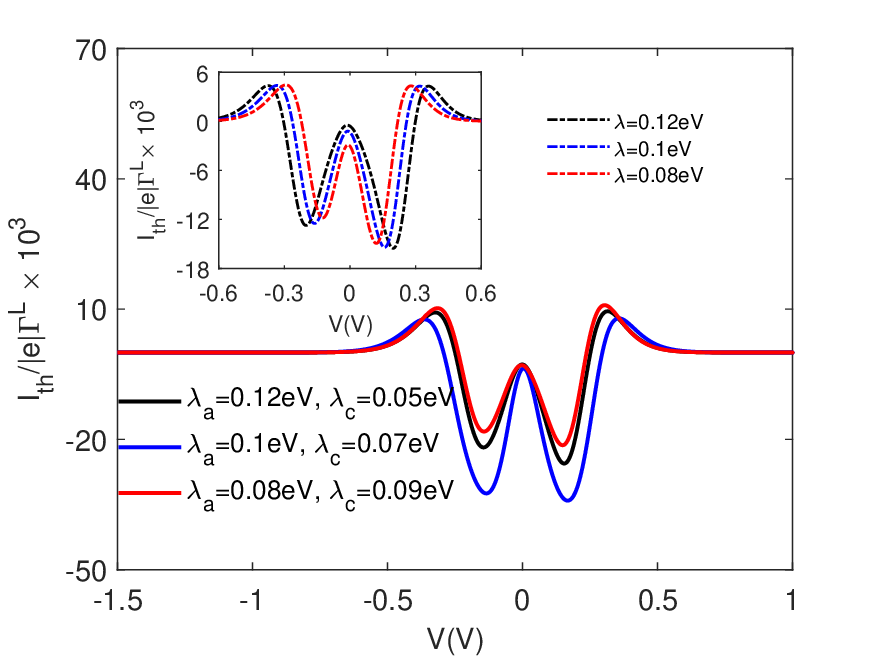} 
\includegraphics[width=4.2cm,height=3.5cm]{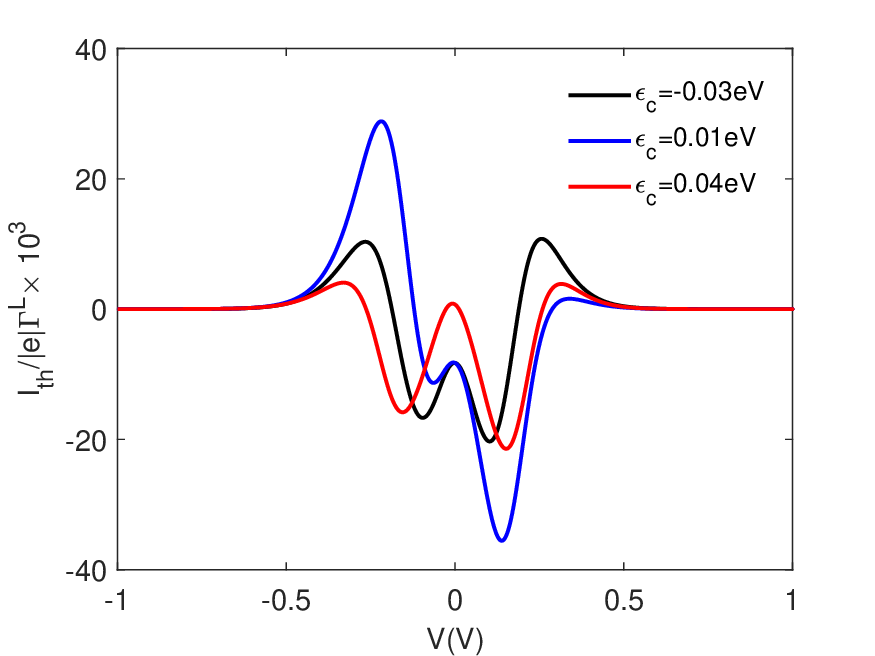}
\caption{The thermocurrent as a function of the bias voltage at the fixed temperature difference $\ds\frac{\Delta T}{T}=0.23$.Curves are plotted assuming that the thermal energy $kT=0.026$eV, $\Gamma=0.01$eV, $\epsilon_a=-0.02$eV, $\epsilon_c=0.03$eV for several values of reorganization energies (left) and assuming that  $\epsilon_a=-0.02$eV, $\lambda_a=0.08$eV, $\lambda_c=0.05$eV for several values of $\epsilon_c$ (right). Inset shows $I_{th}(V)$ plotted using the two states model for the bridge at $\epsilon_a=-0.02$eV and several values of $\lambda$. 
}
 \label{rateI}
\end{center}\end{figure}
Assuming that the bridge is represented by two states, there exist two possible transport channels, and $I_{th}$ changes its direction four times, as shown in the inset in Fig.7. The strengthening of electron interaction with the solvent phonons attested by the increase of the reorganization energy partly shifts the positions of dips in the $I_{th}-V$ and moderately changes the thermocurrent magnitude.

 Simulating the bridge with the three states we may observe diverse patterns of $I_{th}$ behavior. Two examples are displayed in Fig.7. Current -voltage characteristics shown in the left panel are similar to those presented in the inset. This means that only one charged molecular state (in the considered case $|a>$) participates in the process of electron transfer between the electrodes. Transport channels associated with another charged state $|c>$ characterized by a higher energy could open up at stronger bias where the thermocurrent  vanishes. Also, both charged states may simultaneously contribute to the transport at appropriate values of the energies $\epsilon^{\pm}_\alpha$ making the profiles of $I_{th}$ versus $V$ curves more complex, as illustrated in the right panel of the figure. Note that all current-voltage curves shown in Fig.7 are asymmetric with respect to the bias voltage. The asymmetry appears due to the presence of a fixed difference in the electrodes temperatures $\Delta T$. As a result, a thermal driving force emerges which assists or opposes the electric driving force depending on the bias voltage polarity thus causing the asymmetry in $I_{th}-V$ characteristics. Another source for the asymmetry is location of the transport channels. As shown in the figure, the asymmetry is more pronounced  at some values of $\epsilon^{\pm}_{\alpha}$ than at other ones.

\subsection {IV. Conclusions}

In the present work we have theoretically studied steady state thermoelectric properties of metal-molecule junctions assuming that the molecular bridge is put in contact with a dielectric solvent and there exist strong interactions between the solvent phonons and electrons on the bridge molecule. Electron transport is described as sequence of hops between the electrodes and bridge states accompanied by the solvent reorganization. The analysis was carried on basing  on the modified Marcus theory taking into account finite molecular states lifetime. The bridge molecule was simulated by two and/or three states, one of them being neutral and other/others charged with a single electron. The developed analysis leads to the following conclusions.

a) Reorganization processes in the solvent may strongly affect zero bias electron conductance. Even at weak electron -phonon interactions the conductance $G$ decreases whereas the reorganization energy enhances. At stronger coupling of electron to the solvent phonons each peak in the conductance is split in two peaks separated by the energy interval whose width is controlled by the corresponding reorganization energy. Within this interval $G$ remains very close to zero because of Franck-Condon blockade. 

b) At moderate electron-phonon interactions the thermopower $S$ also reduces as a result of reorganization processes in the solvent. Also, it acquires extra changes of sign originating from the splitting of the conduction peaks. On the contrary, at strong electron phonon coupling the thermopower may rise with the increase in the reorganization energy within the regions where the conductance is very low. The detrimental effect of solvent phonons on the electron conductance and thermopower was already discussed \cite{41} but the conductance peaks splitting and corresponding changes in the thermopower were not thoroughly analyzed so far.

c) Whithin the adopted model for the SMJ molecular linker which allows only single electron transitions between the molecular states, the power factor $S^2G$ is strongly reduced by the reorganization processes in the solvent. However, it may be significantly increased if the charged state on the bridge is highly degenerate and several electrons may simultaneously hop to this state from the electrodes thus increasing the conductance \cite{41}.

d) The thermocurrent $I_{th}$ driven by the thermal gradient through an unbiased SMJ emerges when some of energies $\epsilon^{\pm}_\alpha$ are sufficiently close to the chemical potential of electrodes $\mu$. The thermocurrent changes its sign at $\Delta T=0$ and shows linear dependencies on $\Delta T$ at small values of the latter. Reorganization processes in the solvent affect the thermocurrent magnitude. In general, the specifics of the thermocurrent behavior varies following alterations in the energies $\epsilon^{\pm}_\alpha$ which determine how many transport channels stay open at a certain $\Delta T$. 

e) When the SMJ is biased $I_{th}$ is simultaneously driven by thermal and electric forces. It flows through the system at sufficiently low bias where the effects of these forces are comparable and vanishes at higher bias. As in the case of zero bias, the thermocurrent is affected by solvent phonons as well as by the positions of the original (not shifted by electron-phonon interactions) energy levels of the molecular bridge. These factors may strongly change $I_{th}$-$V$ characteristics at a fixed temperature difference between the electrodes.

  The present analysis was carried on assuming that the bridge is equally coupled to the electrodes and the bias voltage is symmetrically distributed. These assumptions do not allow for analysis of possibly significant effects which may appear due to asymmetries in the couplings between the electrodes and the bridge as well as in the  bias distribution over the system. These effects could be a subject of future research. Nevertheless, the present results could be considered as a step in analyzing the effects of reorganization processes in the molecular linker ambience on thermoelectric properties of nanoscale systems. They may appear to be useful in the development of nanoscale technologies which continues to be a focus of research interest.
 
\subsection{Declaration of competing interest}

Authors declare that they have no competing financial interests or personal relationships which could influence the work reported in this paper.

\subsection{Data availability statement}

Data sharing is not applicable as no data are created in this study.

\subsection{Acknowledgments}

The present work was supported by the U.S National Science Foundation (DMR-PREM 2122102).


\begin{thebibliography}{99}

\bibitem{1} Hicks, L. D.; Dresselhaus, M. S. {\it Phys. Rev. B} {\bf  1993},47, 12727.

\bibitem{2} Hicks, L. D.; Dresselhaus, M. S. {\it Phys. Rev. B} {\bf  1993},47, 16631.

\bibitem{3} Aviram, A.; Ratner, M. A. {\it Chem. Phys. Lett.} {\bf 1974}, 29, 277.

\bibitem{4} Agrait, N.; Yeati, A. L.; Ruitenbeck, J.M. {\it Phys. Rep.} {\bf 2003},377, 81.

\bibitem{5} Giazotto, F.; Heikkila, T. T.; Luukanen, A.; Savin, A. M.; Pekkola, J. P. {\it Rev. Mod. Phys..} {\bf 2006}, 78, 217.

\bibitem{6} Cuevas, J.-C.; Sheer, E. {\it Molecular Electronics: An introduction to Theory and Experiment} {\bf 2010}, (World Scientific, Singapur).
.
\bibitem{7} Dubi, Y.; Di Ventra, M. {\it Rev. Mod. Phys.} {\bf 2011}, 83, 131.

\bibitem{8} Zimbovskaya, N. A. {\it J. Phys.: Condens. Matter.} {\bf 2016}, 28, 183002.

\bibitem{9} Tan, A.; Balanchandran, J.; Sadat, S.; Gavini, V.; Dunietz, B. D.; Jang, S.-J.; Reddy, P. {\it J. Am. Chem. Soc.} {\bf 2011}, 133, 8838.

\bibitem{10} Widawski, J. R.; Chen, W.; Vasquez, H.; Kim, T.;Breslow, R.; Hybertsen, M. S.; Venkataraman, l. {\it Nano Lett.} {\bf 2013}, 13, 2889.

\bibitem{11} Gonsanamlou, Z.; Vishkayi,S. T.; Tagani, M. B.; Soleimani, H. R. {\it Chem. Phys. Lett.} {\bf 2014}, 594, 51.

\bibitem{12} Tsutsui, M.; Yokota, K.; Morikawa, T.; Taniguchi, M. {\it Science Rep.} {\bf 2017}, 7, 44276.

\bibitem{13} Kubala, B.; König, J.; Pekkola, J. {\it Phys. Rev. Lett.} {\bf 2008}, 100, 066801.

\bibitem{14} Murphy, P. G.; Mukerjee.; Moore, J. E. {\it Phys. Rev. B} {\bf 2008}, 161406(R).

\bibitem{15} Trocha, P.; Barnas, J. {\it Phys. Rev. B} {\bf 2012}, 85, 085408.

\bibitem{16} Serra, M. A.; Saiz-Brettin, M.; Dominiquez-Adame, F.; Sánchez, D. {\it Phys. Rev. B} {\bf 2016}, 93, 235452.

\bibitem{17} Zimbovskaya, N. A. {\it J. Chem. Phys.} {\bf 2020}, 153, 124712.

\bibitem{18} Wierbicki, M.; Swirkovicz, R.  {\it Phys. Rev. B} {\bf 2011}, 84, 075410.

\bibitem{19} Dubi, Y.; Di Ventra, M. {\it Phys. Rev. B} {\bf 2009}, 79, 081302.

\bibitem{20} Monteros, A. L.; Uppal, G. S.; McMillan, M.; Grisan, M.; Tifrea, I. {\it Euro. Phys. J. B} {\bf 2014}, 87, 302.

\bibitem{21} Arroyo, C. R.; Tarkuc, S.; Frisenda, R.; Seldenthuis, J. S.; Woerde, C. H. M.; Eelkema, R.; Grozema, F. C.; Van der Zant, H. S. J. {\it Angew. Chem. Int. Ed.} {\bf 2013}, 52, 132.

\bibitem{22} Guedon, C. M.; Valkenier, H.; Markussen, T.; Thygessen, K. S.; Hummelen, J. C.; Van der Molen, S. J. {\it Nat. Nanotechnol.} {\bf 2012}, 7, 305.

\bibitem{23} Greenwald,J. E.; Cameron, J.; Findlay, N. J.; Fu, T.; Gunasekaran, S.; Skabara, P. J.; Venkataraman, L. {\it Nature Nanotechnol.} {\bf 2021}, 16, 313.

\bibitem{24} Galperin, M.; Ratner, M. A.; Nitzan, A. {\it J. Phys.:Condens. Matter} {\bf 2007}, 19, 103201.

\bibitem{25} Galperin, M.; Ratner, M. A.; Nitzan, A. {\it Mol. Phys.} {\bf 2008}, 106, 397.

\bibitem{26} Koch, T.; Loos, J.; Feshke, H. {\it Phys. Rev. B} {\bf 2014}, 89, 155133.

\bibitem{27} Härtle, R.; Thoss, M. {\it Phys. Rev. B} {\bf 2011}, 83, 115414.

\bibitem{28} Secker, D.; Wagner, S.; Ballmann, S.; Härtle, R.; Thoss, M. {\it Phys. Rev. Lett.} {\bf 2011}, 106, 136807.

\bibitem{29} Kruchinin, S.; Pruschke, T. {\it Phys. Lett. A} {\bf 2014}, 378, 1157.

\bibitem{30}  Marcus, R. A. {\it J. Chem. Phys.} {\bf 1956}, 24, 966.

\bibitem{31} Marcus, R. A. {\it J. Chem. Phys.} {\bf 1956}, 24, 979.

\bibitem{32} Marcus, R. A. {\it Rev. Mod. Phys.} {\bf 1993}, 65, 599.

\bibitem{33} Migliore, A.; Nitzan, A. {\it ACS Nano} {\bf 2011}, 5, 6669.

\bibitem{34} Migliore, A.; Schiff, P.; Nitzan, A. {\it Phys. Chem. Chem. Phys.} {\bf 2012}, 14, 13746.

\bibitem{35}  Kuznetsov, A. M.; Medvedev, I. G.; Ulstrup, J. {\it J. Chem. Phys.} {\bf 2009}, 131, 164703.

\bibitem{36} Kirchberg, H.; Thorwart, M.; Nitzan, A. {\it J. Phys. Chem. Lett.} {\bf 2020}, 11, 1729.

\bibitem{37} Zimbovskaya, N. A.; Nitzan, A. {\it J. Phys. Chem. B} {\bf 2020}, 124, 2632.

\bibitem{38} Craven, G. T.; Nitzan, A. {\it Proc. Natl. Acad. Sci. USA} {\bf 2016}, 113, 9421.

\bibitem{39} Craven, G. T.; Nitzan, A.  {\it J. Chem. Phys.} {\bf 2017}, 146, 092305.

\bibitem{40} Sowa, J. K.; Mol, J. A.; Andrew, G., Briggs, D.; Gauger, E. M. {\it J. Chem. Phys.} {\bf 2018}, 149, 154112.

\bibitem{41} Sowa, J. K.; Mol, J. A.; Gauger, E. M. {\it J. Phys. Chem. C} {\bf 2019}, 123, 4103.

\bibitem{42} Kirchberg, H.; Nitzan, A. {\it J. Chem. Phys.} {\bf 2022}, 156, 094306.

\bibitem{43} Rincon-Garcia, l.; Evangeli, C.; Rubio-Bollinger, G.; Agrait, N. {\it Chem. Soc. Rev.} {\bf 2016}, 45, 4285.

\bibitem{44} Svensson, S. F.; Hoffmann, E. A,; Nakpathomkin, E. A.; Wu, P. M.; Hu, H. Q.; Nilsson, H. A.; Sánchez, D.; Kashcheyevs, V.; Linke, H. {\it New J. Phys.} {\bf 2013}, 15, 105011.

\bibitem{45} Kim, Y.; Lenert, A.; Meyhofer, E.; Reddy, P. {\it Appl. Phys. Lett.} {\bf 2016}, 109, 033102.

\bibitem{46} Evers, F.; Korytar, R.; Tewari, S.; van Ruiteneck, J. M. {\it Rev. Mod. Phys.} {\bf 2020}, 92, 035001.

\bibitem{47} Cui, L.; Miao, R.; Wang, K.; Thompson, D.; Zotti, L. A.; Cuevas, J. C.; Meyhofer, E.; Reddy, P. {\it Nat. Nanotechnol.} {\bf 2018}, 13, 122.

\bibitem{48} Cui, L.; Miao, R.; Jiang, C.; Meyhofer, E.; Reddy, P. {\it J. Chem. Phys.} {\bf 2017}, 146, 092201.

\bibitem{49} Gehring, P.; Harzheim, A.; Spice, J.; Sheng, Y.; Rodgers, G.; Evangeli, C.; Mishra, A.; Robinson, B. J.; Porfyrakis, K.; Warner, J. H.; Kobosov, O. V.; Briggs, G. A. D.; Mol, J. A. {\it Nano Lett.} {\bf 2017}, 17, 7055.

\bibitem{50} Marko, A.; Cabero, Z.; Guo, Ch.; Wan, c.; Hu, J.; Liu, S.; Zhao, M.; Zhang, L.; Song, Q.; Wang, H.; Tu, S.; Li, N.; Sheng, L.; Chen, J.; Li, Y.; Wei, B.; Zhang, J.; Han, H.; Yu, h.; Yu, D. {\it J. Phys. Chem. C} {\bf 2021}, 125, 13167.

\bibitem{51} Park, S.; Jo, J.-W.; Jang, J.; Ohto, T.; Tada, H.; Yoon, H. J. {\it Nano Lett.} {\bf 2022}, 22, 7682.

\bibitem{52} Wang, R.; Liao, H.; Song, Ch.-Y.; Tang, G.-H.; Yang, N.-X. {\it Sci. Rep.} {\bf 2022}, 12, 12048.

\bibitem{53} Muralidharan, B.; Datta, S. {\it Phys. Rev. B} {\bf 2011}, 83, 115414.

\bibitem{54} Beenakker, C. W. J.; Staring, A. A. M. {\it Phys. Rev. B} {\bf 1992}, 46, 9667.

\bibitem{55} Koch, J.; von Oppen, F. {\it Phys. Rev. Lett.} {\bf 2005}, 94, 206804.

\bibitem{56} Sierra, M. A.; Sánchez, D. {\it Phys. Rev. B} {\bf 2014}, 90, 115313.

\bibitem{57} Sánchez, D.; López, R. {\it Phys. Rev. Lett.} {\bf 2013}, 110, 026804.

\bibitem{58} Meair, J.; Jacquod, P. {\it J. Phys.: Condens. Matter} {bf 2013}, 25, 082201.

\bibitem{59} Whitney, R. R. {\it Phys. Rev. B} {\bf 2012}, 88, 064302.


 \end{thebibliography}
\end{document}